\documentclass[prd,aps,floats,preprintnumbers,preprint]{revtex4}
\newcommand{\sgn}{\mathop{\mathrm{sgn}}}
\usepackage{graphicx,latexsym}
\usepackage{multirow}
 
\newcommand{\be}{\begin{equation}}
\newcommand{\ee}{\end{equation}}
\newcommand{\bea}{\begin{eqnarray}}
\newcommand{\eea}{\end{eqnarray}}
\newcommand{\ba}{\begin{eqnarray}}
\newcommand{\ea}{\end{eqnarray}}

\newcommand{\beq}{\begin{equation}}
\newcommand{\eeq}{\end{equation}}
\newcommand{\beqa}{\begin{align}}
\newcommand{\eeqa}{\end{align}}

\newcommand{\dbar}{{\mathchar'26\mkern-11mu\mathrm{d}}}

\newcommand{\ds}{\mathrm{d}}

\newcommand{\ssl}{\mathrm{s}}

\newcommand{\intfb}[1]{\ensuremath{\int\dbar^4k \,}}
\newcommand{\inttb}[1]{\ensuremath{\int\dbar^3k \,}}

\newcommand{\lp}{\left}
\newcommand{\rp}{\right}

\def\H{{\mathcal H}}
\def\R{{\mathcal R}}

\begin{document}
\setlength{\unitlength}{1mm}
\title{General background conditions for K-bounce and adiabaticity}
\author{Antonio Enea Romano $^{1,2,3}$}

\affiliation{
${}^{1}$Department of Physics, University of Crete, 71003 Heraklion,Greece \\
${}^2$Yukawa Institute for Theoretical Physics, Kyoto University, Kyoto 606-8502, Japan \\
${}^{3}$Instituto de Fisica, Universidad de Antioquia, A.A.1226, Medellin, Colombia \\
}

\begin{abstract}
We study  the background conditions for a bounce uniquely driven by a single scalar field model with a generalized kinetic term $K(X)$, without any additional matter field. 
At the background level we impose the existence of two turning points where the derivative of the Hubble parameter $H$ changes sign and of a bounce point where the Hubble parameter vanishes. 
We find the conditions for $K(X)$ and the potential which ensure the above requirements. 
We then give the examples of two models constructed according to these conditions.  One is based on a quadratic $K(X)$, and the other   a $K(X)$ which is avoiding divergences of the second time derivative of the scalar field, which may otherwise occur. An appropriate choice of the initial conditions can lead to a sequence of consecutive bounces, or oscillations of $H$.
In the region where these models have a constant potential they are  adiabatic on any scale and because of this they may not conserve curvature perturbations on super-horizon scales. While at the perturbation level one class of models is free from ghosts and singularities of the classical equations of motion, in general gradient instabilities are present around the bounce time, because the sign of the squared speed of sound  is opposite to the sign of the  time derivative of $H$. We discuss how this kind of instabilities could be avoided by modifying the Lagrangian by  introducing Galileion terms in order to prevent a negative squared speed of sound around the bounce.

\end{abstract}

\maketitle
\section{Introduction}
The inflationary scenario provides a model able to explain many of the observational features of the observed Universe, such as the cosmic microwave background radiation isotropy or  the spatial flatness. An alternative scenario to solve the horizon problem is the possibility that the Universe underwent a contraction phase before the big-bang, i.e. a bounce. Several attempts have been made to understand if the bounce could be a fully consistent cosmological model \cite{Gordon:2000hv,Khoury:2009my,Khoury:2001wf,Lehners:2008vx,Erickson:2003zm,Khoury:2001bz,Buchbinder:2007ad,ArkaniHamed:2003uy,Creminelli:2006xe,Lehners:2007ac,Creminelli:2007aq,Xue:2011nw,Cai:2012va,Cai:2007qw}, and could indeed be viable alternatives to inflation.

In this paper we study the general conditions to realize the bounce with a K-essence type scalar field, without introducing any additional field. At the background level we impose the existence of two turning points where the derivative of the Hubble parameter $H$ changes sign and of a bounce point where the Hubble parameter vanishes.
We first obtain some general  conditions on the function $K(X)$ and the potential $V(\phi)$ and  then give the examples of two models constructed according to these conditions.  One is based on a quadratic $K(X)$, and the other on  a $K(X)$ which is always avoiding divergences of the second time derivative of the scalar field, which may otherwise occur. 
We integrate numerically the classical equations of motion for these models, verifying that they indeed have the expected evolution, and can produce a bounce.

In the region where these models have a constant potential they became adiabatic on any scale, i.e. globally adiabatic, and because of this they may not conserve curvature perturbations on super-horizon scales. 
We then study the stability of the perturbations, focusingn the attention on the presence of ghosts and gradient instabilities.
We confirm that gradient instabilities  can arise around the bounce because the ghost free condition implies that the sign of the squared speed of sound is opposite to the sign of the time derivative of $H$. Finally we discuss how this kind of instabilities could be avoided by modifying the Lagrangian by  introducing Galileion terms in order to prevent a negative squared speed of sound around the bounce.

\section{$K(X)$-bounce}
We will consider a scalar field with Lagrangian 

\begin{equation}
L_{\phi}=\sqrt{-g}\left[K(X)-V(\phi)\right] \,\,  \,,\,  X=\frac{1}{2}g^{\mu\nu}\partial_{\mu}\phi\partial_{\nu}\phi \label{action} \,.
\end{equation}
We denote the Planck mass with $m_p=(8\pi G)^{-1/2}$, and we adopt adopt units in which $8\pi G/3=1$, so that the Einstein's equation take the form
\bea
H^2=\rho &,& \dot{H}=-\frac{3}{2}(\rho+P)=\frac{3}{2}(K+T)=-3K'X \,, \label{Ht}
\eea
where the energy density and pressure  are 
\bea
\rho&=& 2K' X-K+V=T+V \,,\\
P&=&K-V \,.
\eea
In the above equations we have introduced the function $T(X)$
\bea
T(X)&=&2K'X-K \,,
\eea
because it is convenient to express the equations of motion in terms of it, and it plays a crucial role in defining the conditions for avoiding singularities in the solutions of the classical equations of motion.
The energy conservation equations can then give the equation of motion of the scalar field
\bea
T' \ddot{\phi}+ 3 H  K' \dot{\phi} +V'&=&0 \,. \label{eqphi}
\eea
The minimally coupled scalar field equation of motion  corresponds to the case when $K(X)=T(X)=X$.

Another important quantity to determine the dynamics of the universe is the second derivative of $H$
\bea
\ddot{H}&=&-(3 K' \dot{\phi}\ddot{\phi}+\frac{3}{2} \dot{\phi}^3 K'' \ddot{\phi}) \,,
\eea
which after substituting the equation of motion for $\ddot{\phi}$ gives
\bea
\ddot{H}&=&\frac{3 \left(\dot{\phi}^3 K''+2 \dot{\phi} K'\right) \left(3H \dot{\phi} K'+V'\right)}{2 \left(\dot{\phi}^2 K''+K'\right)} \,.
\eea

We now define \textit{turning points} the local extrema of $H(t)$, where $\dot{H}=0$. From the Einstein's equations we know that $\dot{H}$ can be zero either because $K'=0$ or $X=0$, and the corresponding value of the second order time derivative $\ddot{H}$ will play a crucial role in determining if these are actual extrema or flex points.
From the previous equation after substituting respectively $K'=0$ or $X=0$ we get
\bea
X=0 \, &:& \ \ddot{H}=0 \,,\\
K'=0 \, &:& \ \ddot{H}=\frac{3}{2} V' \dot{\phi}\,.
\eea
This implies that $X=0$ is a flex point and consequently can never be a turning point, while points at which $K'=0$ can be either local maxima or minima depending on the relative sign of $V'$ and $\dot{\phi}$.

\begin{center}
\begin{figure}[h]
\includegraphics[height=60mm,width=150mm]{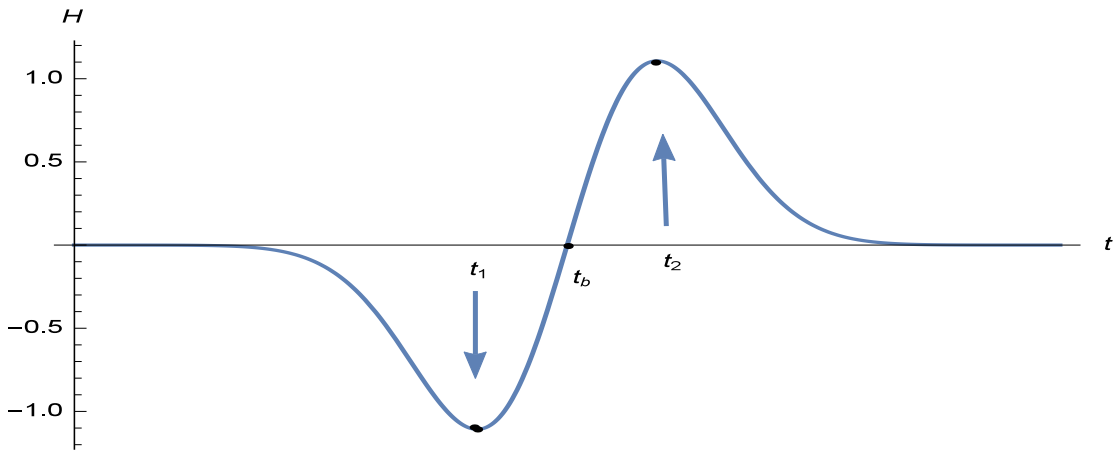}
\caption{Schematic representation of the evolution of $H(t)$ around the bounce. There are two turning points at which the derivative $\dot{H}(t_1)=\dot{H}(t_2)$ is zero, and a bounce point where $H(t_b)=0$}
\label{bounce}
\end{figure}
\end{center}

For a bounce with dynamics similar to the one shown in fig.(\ref{bounce}) to occur, the following three critical points are necessary:
\bea
t_1 \,: \dot{H}=0, \ddot{H}>0 &\rightarrow& K'_1=0\,, V'_1 \dot{\phi}_1>0 \,, \label{BC1}\\ 
t_b \,: \dot{H}>0, H=0 , &\rightarrow& K'_b<0\,,\rho=0 \,, \label{BCb} \\
t_2 \,: \dot{H}=0,\ddot{H}<0 & \rightarrow&  K'_2=0\,,  V'_2 \dot{\phi}_2<0 \label{BC2}\,,
\eea
where we have defined 
\be
K'_i=K'(X(t_i)) \,, V'_i=V'(\phi (t_i)) \,, \dot{\phi}_i=\dot{\phi}(t_i) \,, \label{BC}
\ee
and  $\{t_1,t_2\}$ correspond to the first and second turning point times, while $t_b$ is the bounce time.
The construction  of models satisfying the above conditions will be studied in the following sections.

\subsection{Phase space dynamic of the bounce}
In the previous section we have derived the conditions which have to be satisfied at the turning and bounce points, in eq.(\ref{BC1}-\ref{BC2}). 
In order to give a qualitative description of the phase space behavior of the models is convenient to define $\H^2_0$ as the phase space locus of the points where the energy density is zero, i.e. the set of points satisfying the equation
\bea
3H^2=\rho(X,\phi)=&T(X)+V(\phi)=0 \,.
\eea
The bounce occurs when the phase space trajectory reaches $\H^2_0$, and as shown in appendix A, the trajectory is always tangent to $\H^2_0$. 

The other important points are the ones where $K'=0$, which can correspond to a local minimum or maximum of $H$, depending on the sign of $V' \dot{\phi}$, according to eq.(\ref{BC1}) and eq.(\ref{BC2}). 

It should be noted that even if the trajectory does not reach $\H^2_0$, the conditions for the local extrema of $H$ could be satisfied. This can lead to an oscillating $H$ as shown in fig.(\ref{HV2}), with oscillations that can continue even after the bounce. In phase space this behavior  corresponds to oscillations around the turning point $X_T$ where $K'(X_T)=0$. The conditions in eq.(\ref{BC1}) and eq.(\ref{BC2}) by them self do not guarantee the bounce, but only the existence of local extrema of $H$, which can produce oscillations of $H$. If somewhere between the turning points also the condition in eq.(\ref{BCb}) is satisfied, than there can be a bounce, or a sequence of bounces as shown in fig.(\ref{HV2}). 
The particular choice of $K(X)$, $V(\phi)$ and the initial conditions can determine different phase space behaviors, but we can use the conditions in eq.(\ref{BC1}-\ref{BC2}) in order to design models appropriately.
In order to have turning points we need a kinetic term with at least one local extrema where $K'=0$, while for the choice of the potential there is more freedom as long as $V' \dot{\phi}$ has the right sign at the turning points according to eq.(\ref{BC1}-\ref{BC2}).

A more detailed analytical study of the phase space trajectories and their dependence on the initial conditions goes beyond the scope of the present paper, which is mainly focusing on deriving the general conditions for the bounce and turning points contained in the previous section. We devote the rest of the paper to construct models specifically designed to satisfy those conditions and show by numerical integration of the differential equations that the expected phase space behavior is realized. 
It should be noted that the conditions we obtained in   eq.(\ref{BC1}-\ref{BC2}) are completely general since we do not assume any specific form of $K(X)$  contrary to the analysis performed in other previous works such as \cite{DeSantiago:2012nk,Panda:2014qaa}, where it was assumed $K(X)=X^n$. Another important difference is that we consider models with only one scalar field, without any additional matter field or perfect fluid, and find the most general conditions for a bounce uniquely driven by the scalar field with a dynamics depicted schematically in fig.(\ref{bounce}).

\section{An example: quadratic bounce}
\begin{center}
\begin{figure}[h]
\includegraphics[height=60mm,width=150mm]{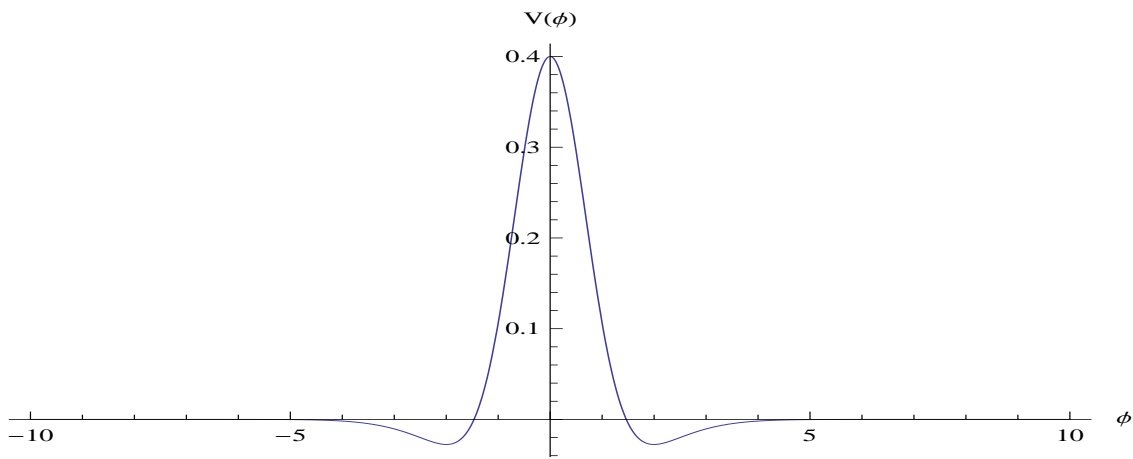}
\caption{The potential for the model in eq.(\ref{V1}) is plotted as a function of the $\phi$ in units of $m_p^2$.}
\label{fig:V1}
\end{figure}
\end{center}

Based on the background conditions derived in the previous section we can introduce a quadratic kinetic term model and an appropriate potential which will give the correct dynamic for $H(t)$:
\bea
K(X)&=&(X-X_T)^2 \,,\\
V(\phi)&=&\frac{1}{2}\left[\alpha_1 e^{\beta_1 x^2}+\alpha_2 e^{\beta_2 x^2}\right] \,.\label{V1}
\eea
In the rest of the paper we will consider a model with $\{\alpha_1=-\beta_1=1,\alpha_2=-0.2,\beta_2=-0.25\}$, which is plotted in fig.(\ref{fig:V1}).

\begin{center}
\begin{figure}[h]
\includegraphics[height=60mm,width=150mm]{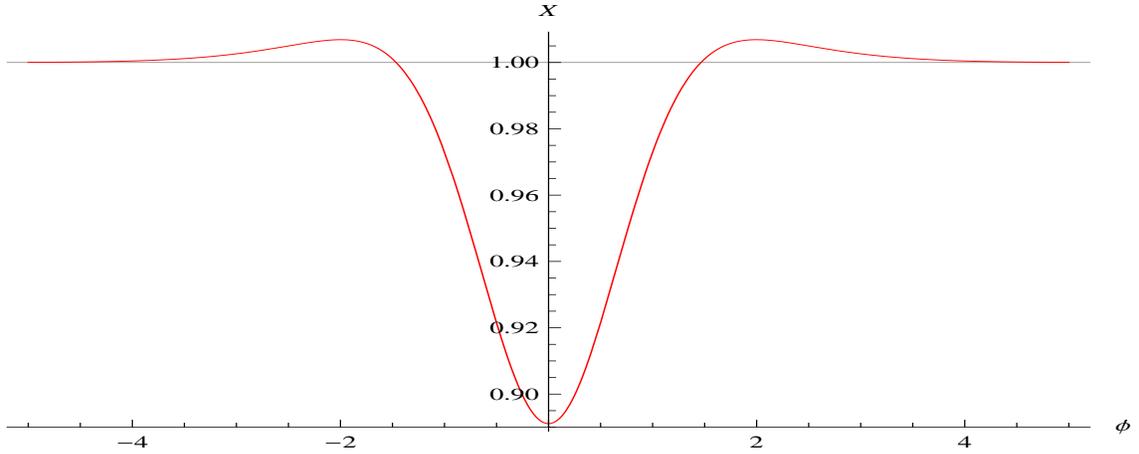}
\caption{The zero energy locus $\H^2_0$ locus is plotted for the model defined in eq.(\ref{V1}). The region below this curve is physically impossible, since it corresponds to $H^2<0$. Physical phase space trajectories can only be above this line, or be tangent to it at the time of the bounce, has proofed in the appendix.}
\label{H2V1}
\end{figure}
\end{center}

\begin{center}
\begin{figure}[h]
\includegraphics[height=60mm,width=150mm]{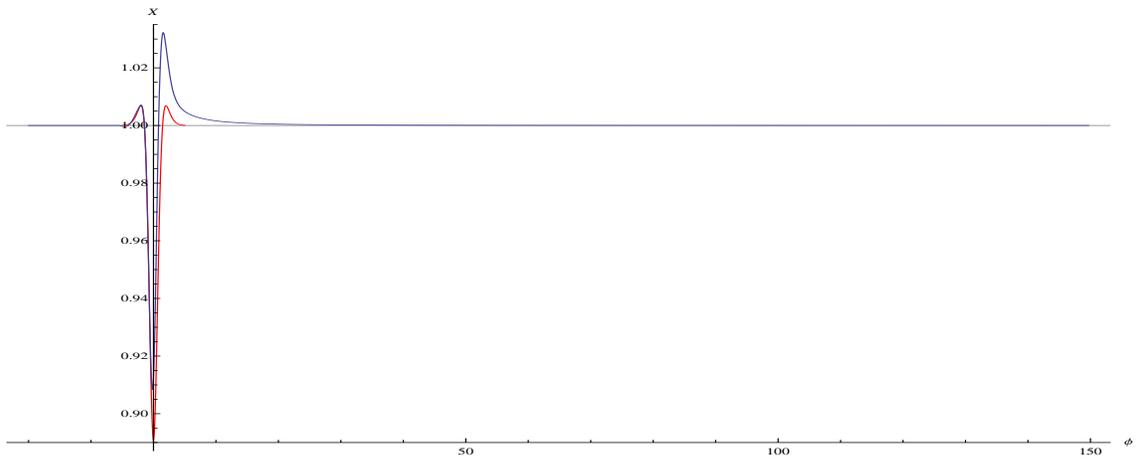}
\caption{The phase space trajectory of the solution of the differential equations for the model defined in eq.(\ref{V1}) is plotted in blue. The red line corresponds to the $\H^2_0$ locus where the energy is zero. As it can be seen the phase space trajectory is tangent to $\H^2_0$ at the bounce time, as proofed in the appendix. The scalar field is plotted in units of $m_p^2$ and $X$ in in units of $m_p^4$.}
\label{Xphi1}
\end{figure}
\end{center}

\begin{center}
\begin{figure}[h]
\includegraphics[height=60mm,width=150mm]{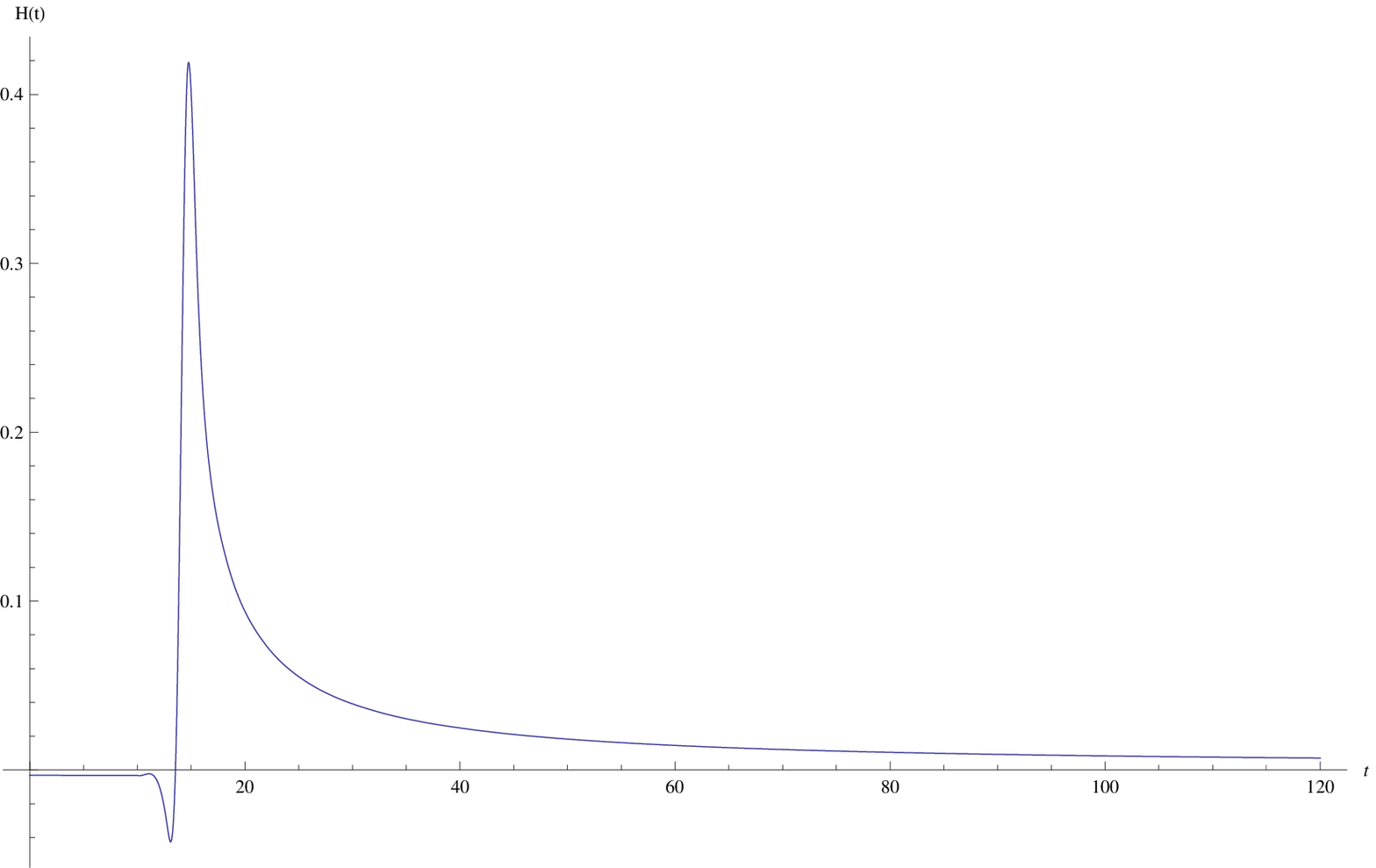}
\caption{H is plotted as function of time  in units of $m_p^{-1}$ for the model defined in eq.(\ref{V1})}
\label{H1}
\end{figure}
\end{center}

\begin{center}
\begin{figure}[h]
\includegraphics[height=60mm,width=150mm]{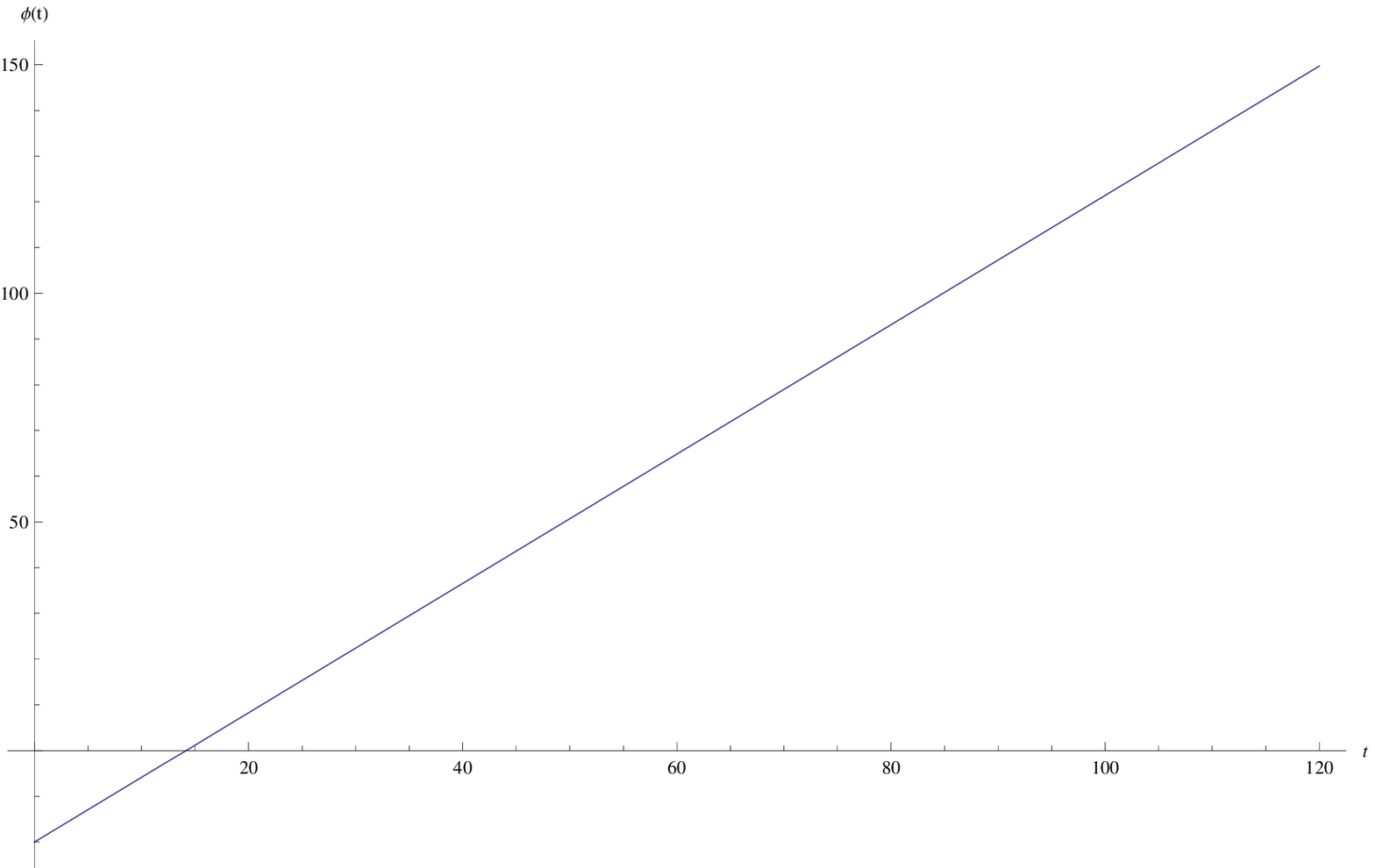}
\caption{The scalar field $\phi$ is plotted as a function of time in units of $m_p^{-1}$ for the model defined in eq.(\ref{V1}).}
\label{phi1}
\end{figure}
\end{center}


\begin{center}
\begin{figure}[h]
\includegraphics[height=60mm,width=150mm]{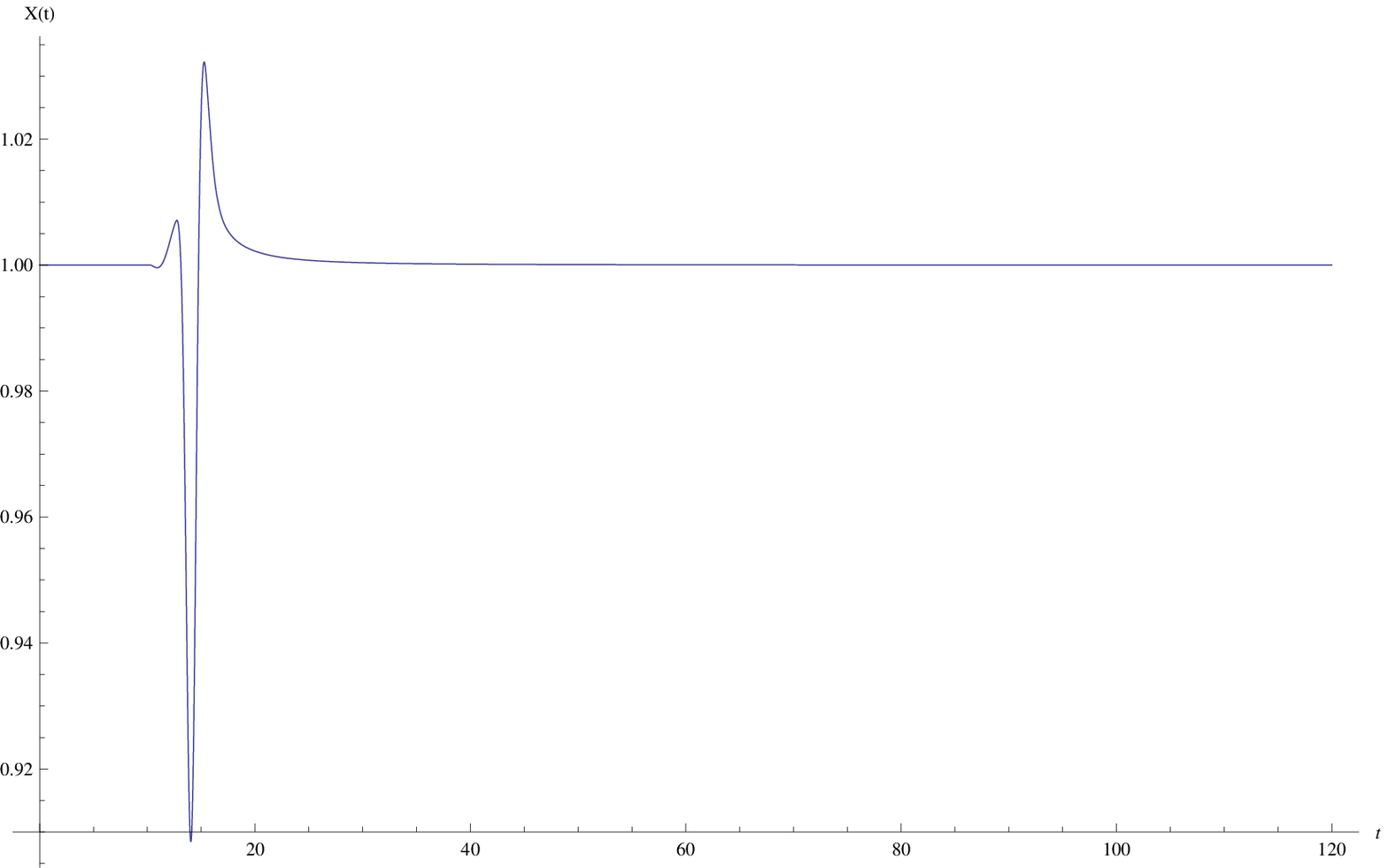}
\caption{$X(t)$ is plotted as a function of time in units of $m_p^2$ for the model defined in eq.(\ref{V1}).  }
\label{X1}
\end{figure}
\end{center}

The phase space behavior of the model defined in eq.(\ref{V1}) is shown in fig.(\ref{fig:V1},\ref{H2V1},\ref{Xphi1},\ref{phi1},\ref{X1}).
As can be seen in fig.(\ref{Xphi1}) the phase space trajectory is tangent to the zero energy locus $\H^2_0$ at the bounce time, as proofed in the appendix.
The turning points correspond to $X=X_T=X(t_1)=X(t_2)$, and the sign of $V'_i \dot{\phi}_i$ determine, according to eq.(\ref{BC1},\ref{BC2}), if the turning point is a local minimum or maximum for $H(t)$.

\section{Possible divergences of the second derivative of $\phi$}
In this section we will try to construct some other model satisfying the conditions in eq.(\ref{BC}). 
Let's consider models of the type
\bea
K(X)&=&(X-X_T)^2 \,,\\
V(\phi)&=&\frac{1}{2} m^2 \phi^2 \,,
\eea
where $X_T$ is the value of $X$ at which the turning point for $H$ should occur, i.e. where $K'(X_T)=0$, based on the considerations in the previous sections. 
Note that in this case we have $X_T=X(t_1)=X(t_2)$, but $V_1\dot{\phi_1}\neq V_2\dot{\phi_2}$ in order to satisfy the conditions for the turning points given in  eq.(\ref{BC1},\ref{BC2}).
The problem with this kind of model is that there is a line in  phase space plane where $T'(X_c)=0$, which corresponds to a divergence of the second derivative of $\phi$. At that point the equation of motion becomes a first order equation, which correspond to a constraint of the value of the field $\phi_c$ when $X=X_c$.
This critical point in phase space is defined by the following conditions:
\bea
T'(X_c)&=& 0 \,,\\
\frac{3 [K(X_c)+T(X_c)] \sqrt{T(X_c)+V(\phi_c)} }{\sqrt{2 X_c}} +V_c'&=&0 \,.
\eea
These  equations can be solved to give
\bea
X_c&=&\frac{X_T}{3} \,,\\
\phi_c&=&-\frac{8 \sqrt{\frac{2}{3}} {X_T}^{5/2}}{\sqrt{-3+16 {X_T}^3}} \,.
\eea
To avoid this kind of divergence we will introduce a new class of models in the next section.

\section{Models avoiding the divergence of the second derivative}
In order to avoid the degenerate solutions encountered in the model studied in the previous section we consider a class of models
for which $T'(X)$ is never zero 
\bea
T'(X)=a \label{Tp} \,,
\eea
where $a$ is a constant.
Solving for $K(X)$ the above differential equation we get
\bea
K(X)&=& a X + b \sqrt{X}+c  \,,
\eea
where $b$ and $c$ are two  arbitrary integration constants. The so called cuscuton model \cite{Afshordi:2006ad} corresponds to the case in which $a=0$.
In this case both $T$ and $T'$ vanish.
In order to obtain a $K(X)$ with a local minimum we need a negative $b<0$, leading to a minimum at
\bea
X_T&=&\frac{b^2}{4} \,.
\eea
Note that an oscillatory $H(t)$ can be obtained for  an appropriate choice of the initial conditions, leading to a sequence of consecutive bounces, or to an oscillating $H(t)$. 

\begin{center}
\begin{figure}[h]
\includegraphics[height=60mm,width=150mm]{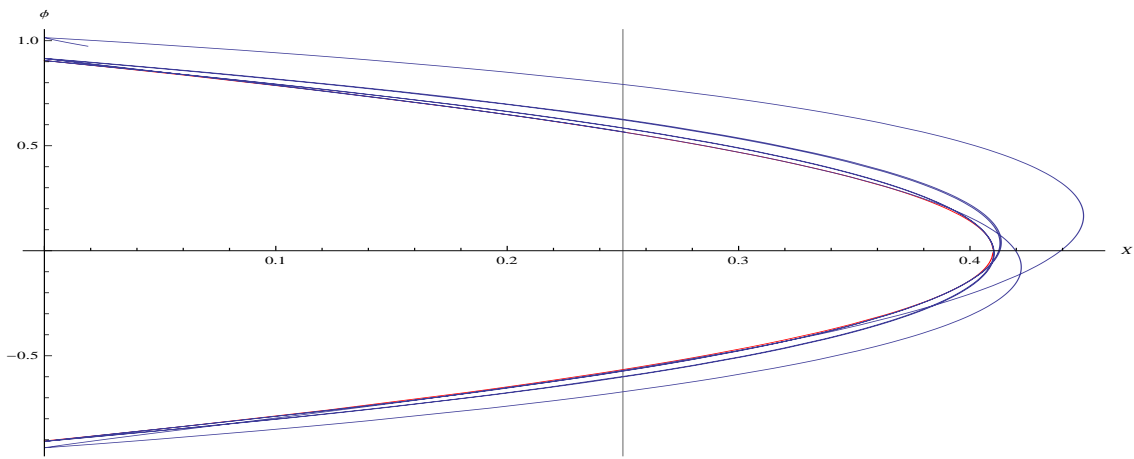}
\caption{Space phase trajectory of the solution of the differential equations for the scalar field $\phi$ is plotted in blue for the model in eq.(\ref{cusB}), with $V_0=-0.41 m_p^4$. The red line corresponds to the zero energy locus  $\H^2_0$. The vertical line is for $X=X_t$, the minimum of $K$ where $K'(X_T)=0$, which corresponds to a turning point for $H(t)$. The points at which the field trajectory is touching the zero energy locus correspond to the bounce times. The scalar field is plotted in units of $m_p^2$ and $X$ in in units of $m_p^4$.} 
\label{Xphi2}
\end{figure}
\end{center}

\begin{center}
\begin{figure}[h]
\includegraphics[height=60mm,width=150mm]{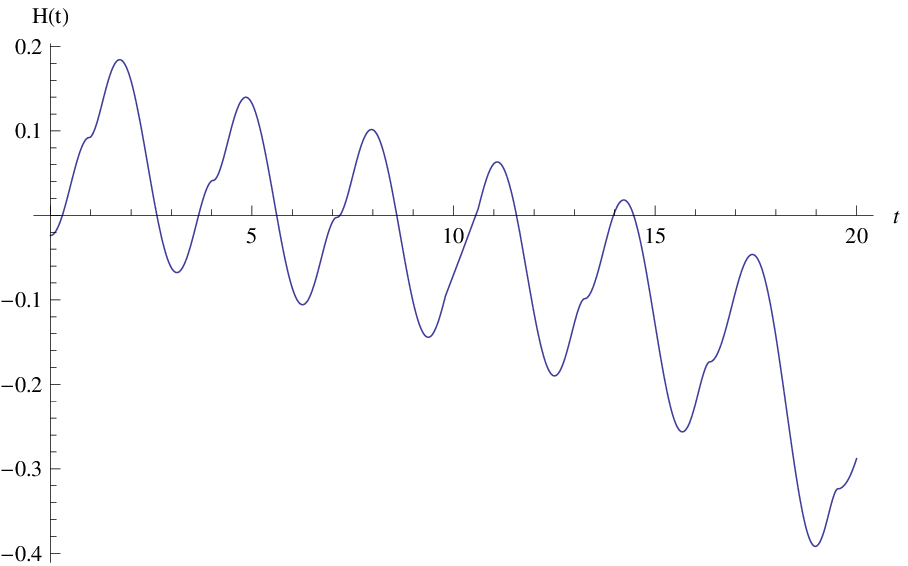}
\caption{H is plotted as function of time  in units of $m_p^{-1}$ for the model defined in eq.(\ref{cusB})}
\label{HV2}
\end{figure}
\end{center}

\begin{center}
\begin{figure}[h]
\includegraphics[height=60mm,width=150mm]{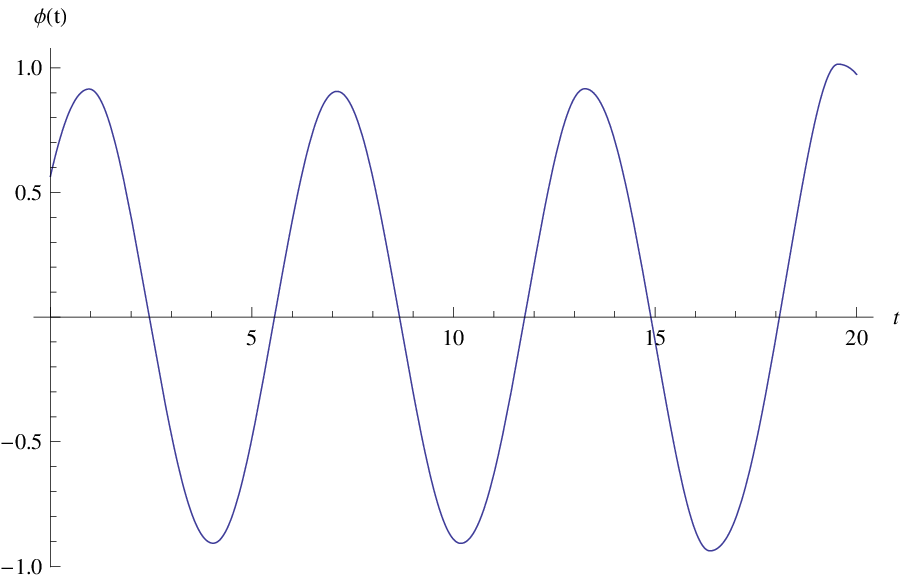}
\caption{The scalar field $\phi$ is plotted as function of time  in units of $m_p^{-1}$ for the model defined in eq.(\ref{cusB}).}
\label{phi2}
\end{figure}
\end{center}

\begin{center}
\begin{figure}[h]
\includegraphics[height=60mm,width=150mm]{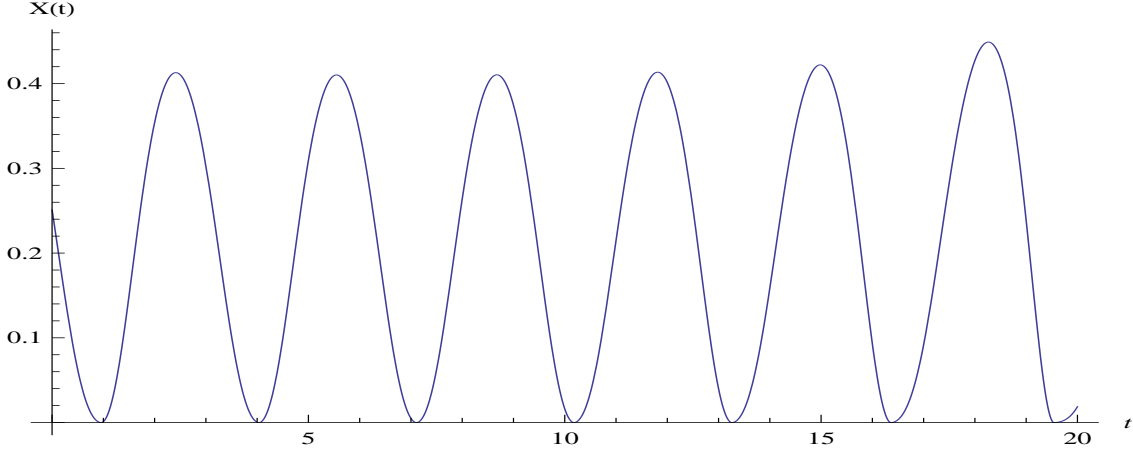}
\caption{$X(t)$ is plotted as a function of time in units of $m_p^4$ for the model defined in eq.(\ref{cusB}).}
\label{X2}
\end{figure}
\end{center}

The linear part of $K(X)$ can give a an ekpyrotic phase, and it arises naturally by imposing the condition $T'(X)=a>0$.
The phase space behavior of the model defined in eq.(\ref{V1}) is shown in fig.(\ref{HV2},\ref{Xphi2},\ref{phi2},\ref{X2}).
As can be seen in fig.(\ref{HV2},\ref{Xphi2}) the phase space trajectory is tangent to the zero energy locus $\H2_0$ at the bounce time, as proofed in the appendix.
As in the case of the model defined in eq.(\ref{V1}) the turning points correspond to $X=X_T=X(t_1)=X(t_2)$, and the sign of $V'_i \dot{\phi}_i$ determine, according to eq.(\ref{BC1},\ref{BC2}), if the turning point is a local minimum or maximum for $H(t)$.
For the purpose of integrating numerically the equations of motion we consider this specific model 
\bea
K(X)&=&X-\sqrt{X} \,,\\ \label{cusB}
K'(X_T)=0 & \,;& X_T=\frac{1}{4} \,, \\
V_2(\phi)&=&V_0+\frac{1}{2}\phi^2 \,.
\eea

\section{Stability of cosmological perturbations and adiabaticity}
So far we have focused on the conditions to realize the bounce at the background level, but in order for these models to be viable cosmological models which could provide an alternative to inflation, we need to check if the cosmological perturbations are stable. In particular there could be problems with the growth of anisotropies \cite{Xue:2011nw} and   gradient instabilities  or ghosts \cite{Deffayet:2010qz}could arise.
The quadratic action for scalar perturbations in the comoving slices gauge for a general Lagrangian of the form $K(X,\phi)+G(\phi) \Box$ can be written as \cite{Deffayet:2010qz}

\beq\label{quadraticaction}
S^{(2)}=\int \ds^3x\ds t a^3 {A\over2} \lp[\dot\zeta^2 - {c_\ssl^2 \over a^2} (\partial_i\zeta)^2  \rp]
~,
\eeq

where we have defined

\bea
A&=&{2XD\over (H-\dot\phi XG_X)^2} \,,\\
D&=&K_X+2XK_{XX}- 2G_\phi - 2XG_{X\phi} +6\dot\phi H(G_X+XG_{XX})+6X^2G_X^2 \,,
\eea
and the squared sound speed for the scalar perturbations is
\begin{equation}\label{cs2_app}
c_\ssl^2 = \frac{K_X-2G_\phi + 2XG_{\phi X} + 2\ddot\phi(G_X+XG_{XX}) + 4\dot\phi HG_X - 2X^2G_X^2}{D} \,.
\end{equation}

The condition for absence of ghosts \cite{Deffayet:2010qz} is $A>0$, while gradient instabilities are avoided when the square of speed of propagation of perturbations $c_s^2$ is always positive. For the class of models we considered, corresponding to $G=0$,  these conditions imply
\bea
D&=&K_X+2XK_{XX}=T'(X) \,,\\
A&=&{2XD\over H^2}>0 \rightarrow T'(X)>0 \,,\\
c_s^2&=&\frac{K_X}{D} >0 \rightarrow  K_X>0 \,,\\
\dot{H}&=&-3K_X(X)X=-3 D X c_s^2 =-3 T'(X) X c_s^2 \,.
\eea
We can see that an additional advantage of the class of models satisfying $T'(X)=a>0$ given in eq.(\ref{cusB}) is that they are ghost free by construction, beside avoiding a singularity of the classical equations since as discussed previously $T'(X)$ is also the coefficient of $\ddot{\phi}$ in eq.(\ref{eqphi}). Nevertheless around the bounce time, when $\dot{a}(t_b)=0$,  gradient instabilities could arise \cite{Vikman:2004dc} because $\dot{a}$ is negative before and positive after $t_b$, which implies $\dot{H}>0$ in some neighborhood of $t_b$. From the Einstein's eq.(\ref{Ht}) we can then immediately deduce that gradient instabilities will necessarily arise in any $K(X)$ model which is also ghost free, i.e. also satisfy $T'(X)>0$, because in this case the sign of $\dot{H}$ is the opposite of the of the sign of $c_s^2$
\be
\sgn{\dot{H}}=-\sgn{c_s^2} \,.
\ee 

These kind of instabilities could be healed  by adding Galileion terms of the form $G(X,\phi)\Box\phi$, to the Lagrangian \cite{Deffayet:2010qz}, because in this case the above relation would not be valid anymore.
Here we can note that a modification consisting in adding terms of the type $G(\phi) \Box$ would not be sufficient
because in this case we have
\ba
\rho&=&2 X(K_X-G_{,\phi})-K \quad; \quad P=K-2 X G_{,\phi} \,, \\
A&=&{2XD\over H^2}>0 \rightarrow D>0 \,, \\
D&=&K_X+2XK_{XX}- 2G_\phi=T'(X)-2G_\phi  \,, \\
c_s^2&=&\frac{K_X-2G_{,\phi}}{D} \,, \\
\dot{H}&=& -\frac{3}{2}(\rho+P) = -3 X(K_X-2G_{,\phi})=-3  D X c_s^2 \,. 
\ea
The last equation implies that also in this case $c_s^2$ must be negative around the bounce, since $\dot{H}$ is positive around the bounce and $D>0$ in order to avoid ghosts. Consequently also for these models, the ghosts free condition implies the relation $\sgn{\dot{H}}=-\sgn{c_s^2}$, and gradient instabilities are unavoidable around the bounce. Only introducing a general $G(X,\phi)$ which is also a function of the kinetic term $X$ the gradient instabilities can be avoided, but even in this case a careful analysis is necessary \cite{Kobayashi:2016xpl,Easson:2011zy,Easson:2016klq}. 

Another general problem of $K(X)$ models related to the sign of $c_s^2$ is that around the bounce the Hamiltonian could be not positive definite when the gradient term dominates the kinetic term. Also this problem could be avoided by adding a Galileion term of the form $G(X,\phi)\Box\phi$ to the Lagrangian in order to avoid a negative sound speed, according to eq.(\ref{cs2_app}), while still realizing the conditions for the bounce.
We will leave to a future upcoming work a detailed study of the stability of cosmological perturbation for $K(X)$ models with the addition of general Galileoion terms, since the purpose of the present paper is mainly  the derivation of general conditions to realize the bounce at the background level with a $K(X)$ model. It should be noted nevertheless that this type of instabilities are general for any $K(X)$ driven bounce model \cite{Vikman:2004dc}, and can arise also for the other models which were  previously studied in \cite{DeSantiago:2012nk,Panda:2014qaa}.

\section{Adiabaticity}

Due to the form of the Lagrangian we can  anticipate that the behavior of curvature perturbations for these models requires a careful treatment. The non-adiabatic pressure perturbation for these kind of scalar field models is in fact zero on any scales \cite{Romano:2016gop} in the region where the potential is constant, because in that regime they are equivalent to a barotropic fluid with equation of state $P(\rho)$ so that
\be
\delta P_{nad}=\delta P-\frac{\dot{P}}{\dot{\rho}}=0 \,,
\ee 
For this reason they are called globally adiabatic(GA) models, to distinguish them from attractor models which are adiabatic only on super-horizon scales (SHA).

After defining the perturbed metric as 
\begin{eqnarray}
ds^2&=&a^2\Bigl[-(1+2A)d\eta^2+2\partial_jB dx^jd\eta+\left\{\delta_{ij}(1+2\R)+2\partial_i\partial_j E\}dx^idx^j\right\}
\Bigr]\,,
\label{metric}
\end{eqnarray}
in \cite{Romano:2015vxz} it was shown that  in general relativity, in absence of anisotropies, the conservation of the perturbed energy momentum tensor gives
\ba
\delta P_{ nad}=
\left[\left(\frac{c_w}{c_s}\right)^2-1\right](\rho+P)\frac{\dot{\R_c}}{H} \,, \label{dpnad} 
\ea
which is true on any scale, and is not based on neglecting any gradient term.
For GA models the adiabatic sound speed $c_s=\left(\frac{\delta P}{\partial \rho}\right)_c$ is equal to the phase speed $c_w=\frac{\dot{P}}{\dot{\rho}}$ and for this reason, despite $\delta P_{nad}=0$, nothing can be inferred about the behavior of $\dot{\R_c}$, otherwise $\R_c$ should be constant on any scale.
Since in ekpyrotic models the potential is not constant during the bouncing phase,  $c_s\neq c_w$. The exponential growth of $\R_c$  \cite{Xue:2011nw} then implies that $\delta P_{ nad}\neq 0$, i.e. the bounce is strongly non adiabatic. This large non adiabaticity is in fact the cause of the growth of anisotropies during the bouncing phase.

On the contrary, in the region where the potential asymptotes a constant value, these models could enter a phase of ultra-slow inflation in which curvature perturbations may not be conserved on super-horizon scales \cite{Romano:2016gop}. This is due to the fact that when the potential is constant $c_s=c_w$  and conservation of curvature perturbation on super-horizon scales is not guaranteed anymore \cite{Romano:2015vxz} despite adiabaticity.
A similar ultra-slow phase could occur for the model defined in eq.(\ref{V1}), in the region where the potential tends to a constant value.
\section{Conclusion}
We have derived the general conditions to obtain a bounce with a scalar field model with generalized kinetic term of the form $K(X)$ in general relativity. The requirement of the existence of a bounce point and of two turning points for $H(t)$ give some conditions which $K(X)$ and the potential $V(\phi)$ have to satisfy.
We have given  the examples of two models constructed according to these conditions.  One is based on a quadratic $K(X)$, and the other on  a $K(X)$ which is always avoiding divergences of the second time derivative of the scalar field, which may otherwise occur. An appropriate choice of the initial conditions can lead to a sequence of consecutive bounces.

While at the perturbation level one class of models is free from ghosts, in general gradient instabilities are present around the bounce time, and they could be avoided only by modifying the Lagrangian. 
In the future it will be interesting to add Galileion type terms to the Lagrangian and study in details the cosmological perturbations during and after the bounce,  and the effects of global adiabaticity in the region where the potential tends to a constant value.

\appendix

\section{Phase space trajectory and zero energy locus}
In general the slope of the tangent at a given point along the trajectory in phase space is
\bea
m&=&\frac{d\phi}{d X}=\frac{\dot{\phi}}{\dot{X}}=\frac{1}{\ddot{\phi}} \,.
\eea
From the equation of motion for $\phi$ we get that at the bounce, when  $H=0$
\bea
\ddot{\phi_b}=-\frac{V'(\phi_b)}{T'(X_b)} \,,
\eea
implying that the phase space trajectory at the time of the bounce has slope
\bea
m_b&=&-\frac{T'(X_b)}{V'(\phi_b)} \,.
\eea
It is convenient to define $\H^2_0$ as the phase space locus of the points where the energy density is zero, i.e. the set of points satisfying the equation
\bea
3H^2=\rho(X,\phi)=&T(X)+V(\phi)=0 \,,
\eea
and then we can get the equation defining $\H_0$ in the form $\phi(X)$ as
\bea
\phi_{\H_0}(X)&=&V^{-1}(-T(X)) \label{phiH0} \,.
\eea 
The slope of the tangent to  $\H_0$ is then given as function of $X$ by
\bea
m_{\H_0}(X)&=&\frac{d\phi_{\H_0}}{dX}=\frac{d V^{-1}(-T)}{d T}\frac{d T}{d X}=-\frac{T'(X)}{V'(V^{-1}(-T))}=-\frac{T'(X)}{V'(\phi_{\H_0}(X))} \,,
\eea
where we have used eq.(\ref{phiH0}) and expressed the derivative of the inverse function $V^{-1}$ in terms if $V'$.

From this we immediately deduce that any scalar field phase space trajectory reaching $\H_0$ at a point $(X_b,\phi_b=\phi_{\H_0}(X_b))$ is tangent to $\H_0$ at that point since
\be
m_b=-\frac{T'(X_b)}{V'(\phi_b)}=-\frac{T'(X_b)}{V'(\phi_{\H_0}(X_b))}=m_{\H_0}(X_b) \,.
\ee
This result is valid for any $K(X)$, and it ensures that only positive energy trajectories are dynamically allowed, since when the scalar field reaches $\H_0$ it is always tangent to it, and never crosses it, which would imply entering the unphysical phase space region where $H^2<0$.

\acknowledgements
I thank Robebert Brandenberger, Yifu Cai and Alexander Vikman for interesting discussions.
This work was supported by the Dedicacion exclusica and Sostenibilidad programs at
UDEA, the UDEA CODI project 2015-4044 and 2016-10945, and Colciencias mobility program.

\end{document}